\documentclass[aps,prb,twocolumn,groupedaddress,floatfix]{revtex4}

\usepackage{graphicx,latexsym,amssymb}

\begin{document}

\newcommand{\fethreeohfour}{F\lowercase{e}$_3$O$_4$}
\newcommand{\fetwoohthree}{F\lowercase{e}$_2$O$_3$}
\newcommand{\fefiveoheight}{F\lowercase{e}$_5$O$_8$}


\title{Surface-Reconstructed Icosahedral Structures for Lead Clusters}



\author{Shaun C. Hendy}
\email[]{s.hendy@irl.cri.nz}
\affiliation{Applied Mathematics, Industrial Research Ltd, Lower Hutt, New Zealand}
\author{Jonathan P. K. Doye}
\affiliation{University Chemistry Laboratory, Lensfield Road, Cambridge CB2 1EW, United Kingdom}


\date{\today}

\begin{abstract}
We describe a family of icosahedral structures for lead clusters. In general,
structures in this family contain a Mackay icosahedral core with a reconstructed
two-shell outer-layer. This family includes the anti-Mackay icosahedra, which have 
have a Mackay icosahedral core but with most of the surface atoms in hexagonal 
close-packed positions. Using a many-body glue potential for lead, we identify 
two icosahedral structures in this family which have the lowest energies of any 
known structure in the size range from 900 to 15000 lead atoms. We show that 
these structures are stabilized by a feature of the many-body glue part of the 
interatomic potential.
\end{abstract}


\maketitle

\section{Introduction}
The structure of an atomic cluster often differs from that of the corresponding bulk material \cite{ino66}.
In such a cluster, the number of surface atoms is comparable to the number of
interior atoms, and consequently, the surface energy plays an important role in determining 
the overall structure. For example, regular noncrystalline structures with fivefold axes of 
symmetry, such as icosahedra and decahedra, are known to occur in gold and a variety of other 
face-centered cubic (fcc) metals \cite{ino66,Allpress67,Marks94}. Such structures are comprised of 
deformed fcc tetrahedral units where adjacent tetrahedral faces meet at a twin plane. The energy 
cost of twinning at the inner tetrahedral faces, and the strain energy in the deformed tetrahedra, 
is overcome by the resulting energetically favorable close-packed outer faces \cite{ino69a,Marks84}.
 
The structure of a cluster, while not only of fundamental interest, is also a key determinant of many of its 
properties. However the delicate balance between surface and internal energies often produces a complex dependence of 
structure upon cluster size \cite{Raoult89,Uzi97}. Eventually, as the size of a cluster increases, the bulk 
structure must win out, but at sizes below this, clusters can assume a variety of regular non-crystalline 
structures. For systems that can be adequately described by pair potentials, there is a relatively good 
understanding of how structure depends on the form of the potential \cite{Doye95,Doye01}. However, for many 
systems of interest, such as metals, the interatomic interactions are more complex. 

Metals exhibit a strong many-body character to their bonding, and because of this, the competition between 
fcc, icosahedral and decahedral structures is less well understood. In addition, many-body effects can potentially 
lead to the emergence of new structural forms \cite{Soler00,Doye02}. In order to study the effect of this bonding on 
structure, it is often necessary to resort to empirical potentials, as {\em ab initio} electronic structure methods are 
prohibitively expensive for all but the smallest cluster sizes. For example, lead clusters have been studied 
using a many-body glue potential \cite{Ercolessi88,Ercolessi91}. The first comprehensive computational study of 
lead cluster structure, by Lim, Ong and Ercolessi \cite{LOE92}, utilized this potential to compare the energetics 
of closed-shell fcc cuboctahedral and Mackay icosahedral structures for cluster sizes from 55 to 3871 atoms. By directly
comparing the binding energies at the same numbers of atoms for each structure, they demonstrated that the fcc cuboctahedra 
were favored over Mackay icosahedra over this size range.

However, electron diffraction of 3-7 nm lead clusters has produced diffraction patterns that cannot be 
adequately fitted by fcc structural models \cite{Hyslop01}. In addition, recent simulations of the melting
and freezing of clusters, using the same glue potential as Lim, Ong and Ercolessi \cite{LOE92}, have 
unexpectedly revealed that fcc structures were not the lowest in energy in this size range\cite{Hendy01}. In 
these simulations, clusters were prepared by the quenching of liquid lead droplets. This procedure was found to produce 
icosahedra overwhelmingly, and these resolidified icosahedra were found to be energetically favored over fcc 
structures \cite{LOE92}. The resolidified icosahedra resembled anti-Mackay icosahedra \cite{Doye97} which 
have a Mackay icosahedral core but with most of the outer layer in hexagonal close-packed (hcp) surface 
sites. The improved stability of these icosahedra was evidently due to this surface reconstruction.

The purpose of this work is to examine the surface reconstruction of these resolidified icosahedra in more 
detail. In particular, we will show that the surface reconstruction is similar, but not identical, to that 
of the anti-Mackay icosahedra. These surfaces features lead us to identify a new family of icosahedral structures,
of which the anti-Mackay icosahedron is a member. We will describe these structures in detail, and show how they 
can lead to lower energy structures for the lead glue potential. 

\section{Computational Techniques}

\subsection{Molecular dynamics}

Molecular dynamics was performed using a local version of the classical molecular dynamics code
ALCMD, originally developed by Ames Laboratory. Finite temperature simulations were performed in 
the microcanonical ensemble (i.e. constant energy). The time step was chosen as $3.75 \ \mbox{fs}$ 
throughout. Melting and freezing simulations were carried out using the procedure detailed in Hendy
and Hall\cite{Hendy01}. 

The inter-atomic potential used is due to Lim, Ong and Ercolessi \cite{LOE92}. This is a
many-body glue-type potential, given by
\begin{eqnarray}
E & = & E_{\mbox{\scriptsize pair}} + E_{\mbox{\scriptsize glue}}, \nonumber \\
& = & \sum_{i < j} \phi \left( r_{ij} \right) + \sum_{i} U(n_i) \; , \label{glue}
\end{eqnarray}
where $\phi$ is a short-range pair potential and $U(n)$ is a many-body glue term which
reflects the effects of non-local metallic bonding. The quantity $n_i$ is a ``generalized 
coordination number'' for atom $i$ defined as:
\begin{equation}
\label{coord}
n_i=\sum_{j} \rho(r_{ij})
\end{equation}
where $\rho$ is some short-ranged ``atomic density'' function. In practice, the function $\rho(r_{ij})$ has a 
cut-off $r_{\mbox{\scriptsize cut}}$, beyond which $\rho(r_{ij})=0$, and for the potential here, a value of 
$r_{\mbox{\scriptsize cut}} = 5.503 \mbox{\AA}$ is used. This cut-off typically lies between the second and third 
neighbor shells. The three functions $\phi$, $\rho$ and $U$ have been obtained by fitting to a number of 
known properties of lead including cohesive energy, surface energy, elastic constants, phonon frequencies, 
thermal expansion and melting temperature \cite{Ercolessi88}. This potential has been used previously to 
model lead clusters \cite{LOE92,LOE94,Hendy01}, temperature-dependent surface reconstructions of low-index 
lead surfaces \cite{TOE94} and pre-melting of low-index lead surfaces \cite{Uzi95}. Recently, putative global 
minimum energy cluster structures were determined for this potential, for cluster sizes of up to 160 
atoms \cite{Doye02}. 

\subsection{Common neighbor analysis}

Common neighbor analysis \cite{CNA93} (CNA) has been used here to analyze cluster structures \cite{Uzi97,Uzi99}. 
CNA is a decomposition of the radial distribution function (RDF) according to the local environment of each
pair. We consider that the first peak in the RDF represents ``bonded" neighbors. As such, if $r_c$
is the first minimum in the RDF, we classify any pair separated by $r < r_c$
as a bonded pair. With this identification, any pair contributing to the RDF can be classified by a set
of three indices, $ijk$, which provide information on the local environment of the pair. The first
index, $i$, is the number of bonded neighbors common to both atoms. The second index, $j$, is the number
of bonds between this set of common neighbors. The third index, $k$, is the number of bonds in the longest
continuous chain formed by the $j$ bonds between common neighbors. 

CNA is useful here because it allows one to distinguish between local atomic arrangements, including fcc 
and icosahedral environments, using the type and number of $ijk$ indices of each atom. For each atom, we can define 
$n_{ijk}$ to be the number of bonds of this atom with CNA indices $ijk$. We can then classify the local
environment of each atom using these $n_{ijk}$ values. In Table~\ref{CNA} we have listed the classifications 
of CNA signatures used here to label the local environment of an atom (this classification is similar but 
not identical to that used by Cleveland, Luedtke and Landman\cite{Uzi99}). We note that these signatures are 
based only on the CNA decomposition of the first peak in the RDF. 

\begin{table*}
\caption{Description of CNA signatures used.}
\label{CNA}
\begin{ruledtabular}
\begin{tabular}{clll}
label & description & position & classification of pairs \\
\hline
A & fcc internal atom & internal & $n_{421} \geq 4$  \\
B & fcc \{111\} face atom & surface & $n_{311} \geq 3$ and $n_{322}= 0$\\
C & fcc \{100\} face atom & surface & $n_{211} \geq 3$ \\
D & fcc \{111\}/\{100\} edge atom & surface & $n_{211} = 2$ and $n_{311} = 2$ \\
E & internal atom at a \{111\} fcc stacking fault & internal & $n_{422} \geq 5$ \\
F & internal icosahedral atom (spine or central atom) & internal & $n_{555} \geq 2$ \\
G & surface icosahedral apex atom & surface & $n_{555} \geq 1$ \\
H & surface icosahedral \{111\}/\{111\} edge atom & surface & $n_{311} \geq 3$, $n_{322} \geq 1$ and $n_{322} \geq 1$ \\
I & anti-Mackay surface \{111\}/\{111\} edge atom & surface & $n_{311} = 2$, $n_{200} = 2$ and $n_{211} \ge 1$\\
? & unclassified signature (possibly disordered) & internal/surface &\\
\end{tabular}
\end{ruledtabular}
\end{table*}

\section{Anti-Mackay Icosahedra}

\begin{figure}
\resizebox{\columnwidth}{!}{\includegraphics{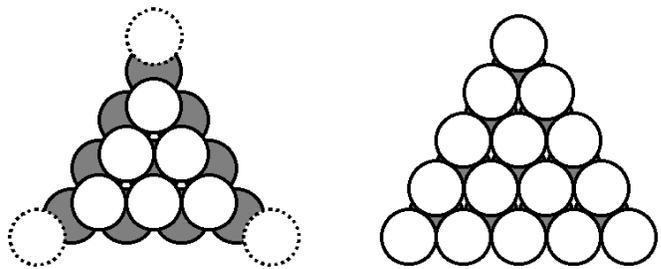}}
\caption{The anti-Mackay surface termination (left) has surface atoms in
hcp positions. The surface terminations considered here do not include the 
vertex atoms (indicated by the dotted lines). The Mackay surface termination 
(right) is shown for comparison.}
\label{anti}
\end{figure}

We will refer to an icosahedron with an anti-Mackay surface termination as an anti-Mackay icosahedron. With 
such a surface termination, atoms in the exterior shell lie in hcp positions relative to a core Mackay 
icosahedron, as illustrated in figure~\ref{anti}. Note that for the terminations we will consider here,
we neglect the icosahedral vertex atoms on the surface since these are not present in the resolidified 
icosahedra, and tend to increase the overall energy of the cluster.

An $n$-shell anti-Mackay icosahedron contains a core $(n-1)$-shell Mackay icosahedron with $1/2 (n-3)(n-4)+3(n-2)$ 
surface atoms per face (neglecting the 12 vertices). The total number of atoms in an $n$-shell Mackay icosahedron 
is
\begin{equation}
\mbox{Ico}(n) = {10 \over 3}n^3+5n^2+{11 \over 3}n+1,
\end{equation}
giving the sequence 55, 147, 309, 561, 923, 1415, 2057, 2869, 3871 ...
Hence, the total number of atoms in an $n$-shell anti-Mackay icosahedron is
\begin{equation}
\mbox{Anti}(n) = {10 \over 3} n^3 + 5 n^2 - {19 \over 3} n - 1  
\end{equation}
This gives a sequence of closed-shell anti-Mackay icosahedra with numbers of atoms as 
follows: 115, 267, 509, 861, 1343, 1975, 2777, 3769, 4971, 6403, 8085 {\ldots} 

Thus, the n-shell anti-Mackay icosahedron contains $10n+2$ fewer surface atoms than the n-shell Mackay icosahedron, 
so the packing of the surface atoms will be less dense. However, these missing atoms come from edges and vertices with
low coordination $n_i$. Table~\ref{compare} compares the energy of an 8-shell Mackay icosahedron and an 8-shell anti-Mackay 
icosahedron, broken down by CNA label. Here we can see that while the surface binding energy per atom is worse
for the anti-Mackay icosahedron, the binding energy per atom for the cluster as a whole is better than the Mackay 
icosahedron. Thus for the potential (\ref{glue}), this removal of edges and vertices from the Mackay icosahedron 
(which can be seen to have particularly poor energetics in Table~\ref{compare}), improves the total energy per atom. 
\begin{table*}
\caption{Comparison of average energies of a closed-shell Mackay icosahedron, a closed-shell
anti-Mackay icosahedron, a resolidified icosahedron and a $m=7$ surface-reconstructed icosahedron.}
\label{compare} 
\begin{ruledtabular}
\begin{tabular}{l|c|c|c|c|c|c|c|c}
& \multicolumn{8}{c}{Cluster} \\ \hline & \multicolumn{2}{c}{Mackay} & \multicolumn{2}{c}{Anti-Mackay}& \multicolumn{2}{c}{Resolidified} & \multicolumn{2}{c}{$m=7$} \\ 
 & \multicolumn{2}{c}{Icosahedra} & \multicolumn{2}{c}{Icosahedra}& \multicolumn{2}{c}{Icosahedra}& \multicolumn{2}{c}{Icosahedra} \\ 
Signature & N & $\left< E \right>$ & N & $\left< E \right>$ & N & $\left< E \right>$ & N & $\left< E \right>$\\  
& & (eV/atom) & & (eV/atom)& & (eV/atom) & & (eV/atom) \\ 
\hline & &  & &  & &  & &\\ 
A (fcc) & 700 & -2.0159 & 400 & -2.0264 & 468 & -2.0048 & 467 & -2.0062 \\ 
B (111) & 420 & -1.6254 & 200 & -1.5836 & 173 & -1.6963 & 185 & -1.6780 \\ 
C (100) & -  & - & 1 & -1.4007 & 75 & -1.6425 & 82 & -1.6254 \\ 
D (fcc edge) & - & - & - & - & 2 & -1.2624 & - & - \\ 
E (hcp) & 630 & -2.0103 & 750 & -2.0071 & 654 & -2.0082 & 646 & -2.0083 \\ 
F (ico) & 85 & -1.9916 & 265 & -1.9794 & 183 & -2.0016 & 195 & -1.9990  \\ 
G (surf ico) & 12 & -1.0555 & - & - & 15 & -1.7832 & 1 & -1.7702  \\ 
H (ico edge) & 210 & -1.4633 & - & - & 1 & -1.9126 & - & - \\ 
I (anti edge) &  - & - & 298 & -1.5531 & 230 & -1.6301 & 276 & -1.6212 \\ 
? surface & - & - &  61 & -1.3585 & 137 & -1.5031 & 83 & -1.6652  \\ 
? interior & - & - & - & - & 119 & -1.9810 & 138 & -1.9811 \\
\hline
surface & 642 & -1.5617 & 560 & -1.5425 & 633 & -1.6251 & 627 & -1.6181  \\ 
bulk & 1415 & -2.0120 & 1415 & -2.0074 & 1424 & -2.0040 & 1446 & -2.0038 \\
total & 2057 & -1.8714 & 1975 & -1.8755 & 2057 & -1.8874 & 2073 & -1.8871 \\
\end{tabular}
\end{ruledtabular}
\end{table*}

\begin{figure}
\resizebox{\columnwidth}{!}{\includegraphics{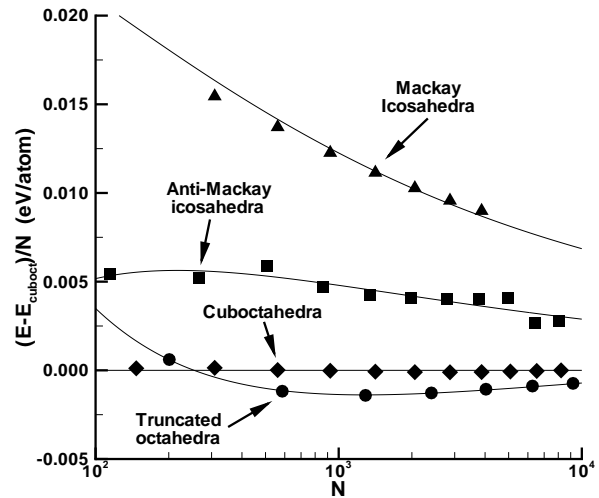}}
\caption{Energies of clusters versus size: anti-Mackay icosahedra (solid square), cuboctahedra (solid diamond), 
Mackay icosahedra (solid triangle) and truncated octahedra (solid sphere). The energies are given relative
to a fit to the energies of the cuboctahedra sequence: $E_{\mbox{\scriptsize cuboct}}=-2.0293 N + 1.8216 N^{2/3} 
+ 0.7134 N^{1/3}$.}
\label{anti-energy}
\end{figure}

However, despite this improvement in the binding energy of the anti-Mackay icosahedra, they are still not 
energetically competitive with the cuboctahedra sequences or the resolidified icosahedra, as shown in 
figure~\ref{anti-energy} (note that here we refer to cuboctahedra with triangular (111)-faces simply as 
cuboctahedra, and to cuboctahedra with hexagonal (111)-faces as truncated octahedra). Thus, while the 
resolidified icosahedra and the anti-Mackay icosahedra share a similar surface reconstruction, the 
resolidified icosahedra have other features which account for their more favorable energetics. We will 
discuss these features in the next section.

\section{Resolidified Icosahedra}

\begin{figure}
\resizebox{\columnwidth}{!}{\includegraphics{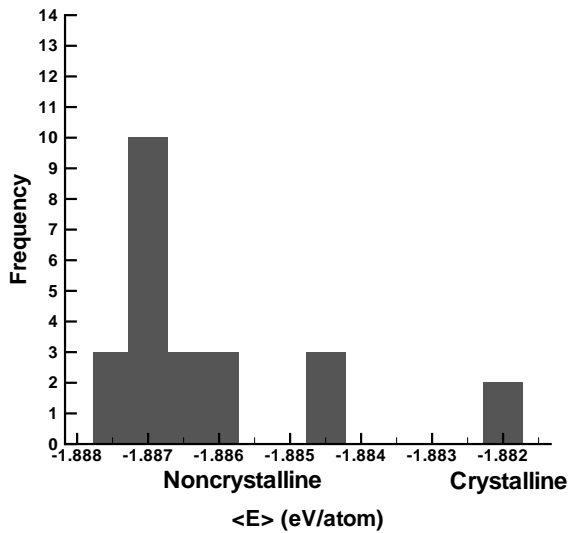}}
\caption{Histogram showing the distribution of energies ($\left<E\right>$) for 25 resolidified 2057 atom
clusters. Two resolidified clusters were identified as fcc truncated 
octahedra (with energies of approximately -1.8820 eV/atom) while the remaining 23 were identified
as icosahedron-like (energies below -1.8840 eV/atom).}
\label{solidtrial}
\end{figure}

\begin{figure}
\resizebox{\columnwidth}{!}{\includegraphics{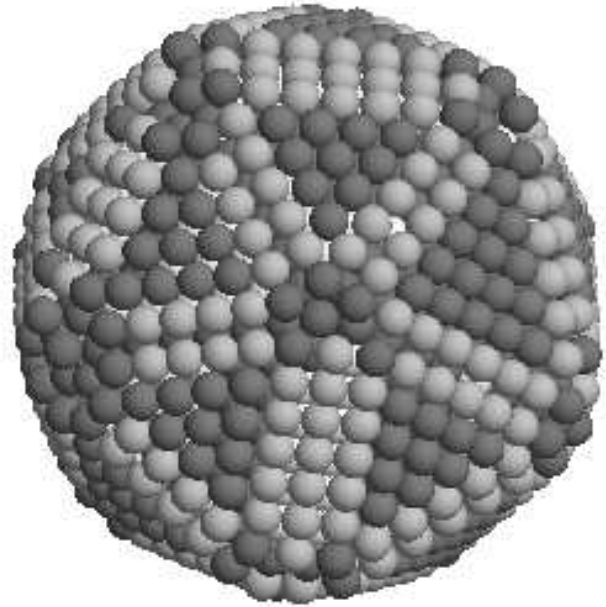}}
\caption{A 2057-atom resolidified icosahedra. Note the anti-Mackay type
surface reconstruction but also the extra \{100\}-facet. This structure had the
highest binding energy of any cluster produced in the resolidification trials.}
\label{iconewxyz}
\end{figure}

Hendy and Hall \cite{Hendy01} conducted a series of resolidification trials, 
where lead clusters were melted and then resolidified at constant energy. 
Figure~\ref{solidtrial} shows the distribution of binding energies of 2057-atom clusters 
that emerged from a typical sequence of 25 resolidification trials (more details can be found in 
Hendy and Hall \cite{Hendy01}). These trials typically produce
icosahedron-like structures, similar to that shown in figure~\ref{iconewxyz}, which have higher 
binding energies than comparably-sized fcc clusters such as cuboctahedra or 
truncated octahedra. In fact, the cluster shown in figure~\ref{iconewxyz}
had the highest binding energy produced in the trial. Hendy and Hall \cite{Hendy01} noted 
that these icosahedron-like clusters had similar surface reconstructions to the 
anti-Mackay clusters.

In table~\ref{compare}, we compare the structures of an anti-Mackay icosahedron 
and a resolidified icosahedron of a similar size. A crucial difference is the denser 
surface packing of the resolidified icosahedron. The number of atoms on the surface 
is comparable to that of the Mackay icosahedron (see table~\ref{compare}), without 
the inclusion of energetically unfavorable edge atoms. Instead, there are a number of 
extra \{100\}-facets (see figure~\ref{iconewxyz}) distributed about the surface. 
Consequently, the binding energy of the surface atoms of the resolidified icosahedron 
is substantially larger than that in both the Mackay and anti-Mackay icosahedra.

\begin{figure}
\resizebox{\columnwidth}{!}{\includegraphics{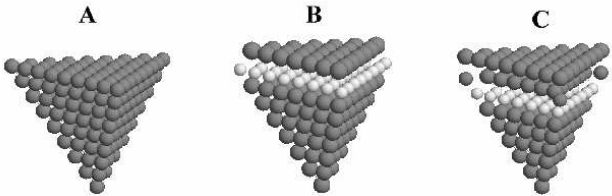}}
\caption{Tetrahedron with a regular Mackay surface termination (A), an anti-Mackay surface termination with a 
stacking fault in the penultimate layer (B), and tetrahedron with a twin plane one layer lower (C).}
\label{icotet}
\end{figure}

A closer examination of the resolidified icosahedra reveals how they differ from anti-Mackay icosahedra.
Recall that a Mackay icosahedron can be constructed from twenty fcc tetrahedra. Likewise, an anti-Mackay icosahedron 
can be constructed from twenty tetrahedra, each of which has a surface with atoms in hcp positions, as 
shown in figure~\ref{icotet}. We refer to the tetrahedra that make up a Mackay icosahedron as type A. The tetrahedra 
that make up an anti-Mackay icosahedron will be referred to as type B. Note that the type B tetrahedra has a stacking 
fault in the penultimate layer (...ABCA\underline{B}A) as the surface atoms lie in hcp positions. The resolidified icosahedra 
have been found to consist of a mixture of type B tetrahedra, and a third type of tetrahedra, which we will refer to as type C. 
Type C tetrahedra have a twin plane in the third shell from the surface (...ABC\underline{A}CB). This third type of tetrahedra 
is also shown in figure~\ref{icotet}.

\begin{figure}
\resizebox{\columnwidth}{!}{\includegraphics{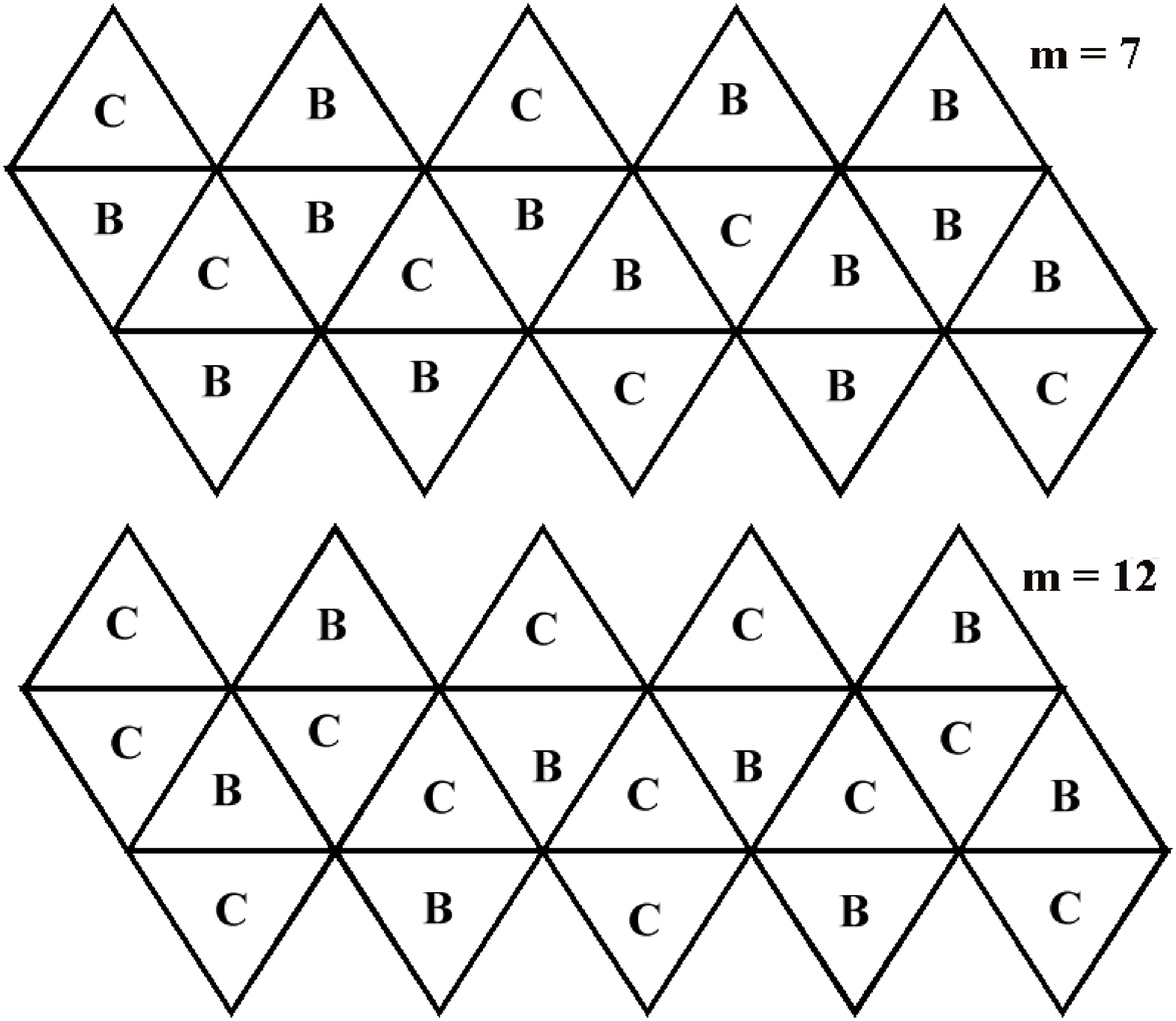}}
\caption{This figure shows two arrangements of type-B and type-C tetrahedra considered here. 
The top pattern (with $m=7$ type-C tetrahedra) shows the arrangement of tetrahedra in the
2057-atom resolidified icosahedron shown in figure~\ref{iconewxyz}. The lower pattern (with $m=12$
type-C tetrahedra) shows the arrangement of tetrahedra in the icosahedra which appear to be stable at 
larger sizes.}
\label{ico-design}
\end{figure}

The arrangement of these type B and type C tetrahedra for the 2057-atom resolidifed cluster from 
figure~\ref{iconewxyz}, is shown in figure~\ref{ico-design}. The extra \{100\}-facets, visible in figure~\ref{iconewxyz}, 
occur at some of the edges between type B and type C tetrahedra. Table~\ref{compare} compares the energies of the 
2057-atom resolidified icosahedron, cuboctahedron and Mackay icosahedron. It is clear that this surface reconstruction 
considerably lowers the surface energy of the resolidified icosahedra. Overall, the 2057-atom resolidified icosahedron 
has a total energy per atom that is 5 meV lower than the cuboctahedron, and 16 meV lower than the Mackay icosahedron.

\begin{table*}
\caption{A comparison of the surface, bulk and total energies per atom of a 2057-atom cuboctahedron, 
a 2057-atom closed-shell anti-Mackay icosahedron and the 2057-atom resolidified icosahedron 
in figure~\ref{iconewxyz}. Also shown is the decomposition of the total energy into 
$E_{\mbox{\scriptsize pair}} + E_{\mbox{\scriptsize glue}}$ (equation~\ref{glue}), and the strain 
energy $E_{\mbox{\scriptsize strain}}$ (equation~\ref{strain}).} 
\label{table_surf}
\begin{ruledtabular}
\begin{tabular}{c|ccccccc}
Cluster 
& $\left<E_{\mbox{\scriptsize bulk}}\right>$ 
& $\left<E_{\mbox{\scriptsize surface}}\right>$ 
& $\left< E \right> $
& $\left<E_{\mbox{\scriptsize glue}}\right>$ 
& $\left<E_{\mbox{\scriptsize pair}}\right>$ 
& $\left<E_{\mbox{\scriptsize strain}}\right>$
& $n_{nn}$ \\
& \multicolumn{6}{c}{(eV/atom)} \\
\hline
Cuboctahedron & -2.0208  & -1.5901  & -1.8817 & -1.7368 & -0.14492 & 0.01609 &  11040\\
Mackay Ico &  -2.0141  & -1.5510 & -1.8714  & -1.7253 & -0.14614 & 0.01802 &  11256\\
Resolidified Ico & -2.0032  & -1.6275 & -1.8876  & -1.7522 & -0.13544 & 0.02634 & 11093\\
\end{tabular}
\end{ruledtabular}
\end{table*}

Table~\ref{table_surf} further decomposes the total energy of 2057-atom structures into the glue 
and pair potential components. The resolidified icosahedron is able to achieve a considerably lower 
glue energy than the two conventional structures, which more than compensates for an increase 
in the pair energy. We can further decompose the pair energy into two parts \cite{Doye95}:
\begin{equation}
\label{strain}
E_{\mbox{\scriptsize pair}} = -n_{nn} \epsilon + E_{\mbox{\scriptsize strain}}
\end{equation}
where $n_{nn}$ is the number of nearest neighbors, $\epsilon$ is the depth of the pair potential
and $E_{\mbox{\scriptsize strain}}$ is the energetic penalty for pair distances that deviate from 
$r_{\mbox{\scriptsize min}}$, the position of the minimum of the pair potential $\phi$. Table~\ref{table_surf} 
shows that the increase in the pair energy of the resolidified icosahedra comes chiefly from an 
increase in the strain energy. Thus, the surface reconstruction of the resolidified icosahedra is 
able to considerably improve the surface energy via the glue term, incurring a smaller increase 
in strain energy. 

\begin{figure}
\resizebox{\columnwidth}{!}{\includegraphics{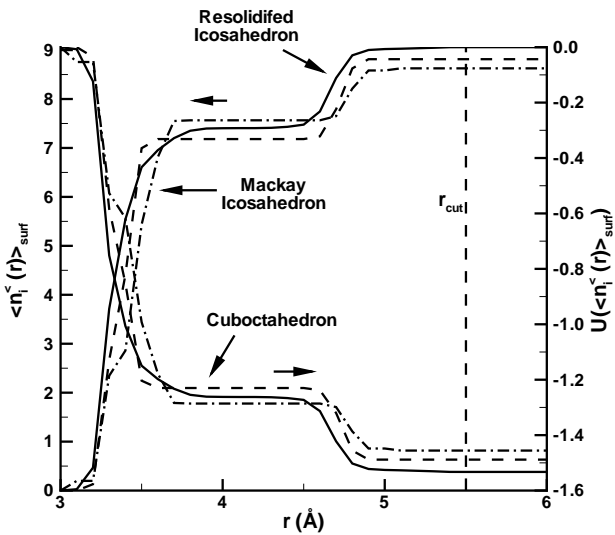}}
\caption{A comparison of $\left< n_i^< (r)\right>_{\mbox{\scriptsize surf}}$ and 
$U\left(\left< n_i^< (r)\right>_{\mbox{\scriptsize surf}} \right)$ for the $m=7$ 2057-atom resolidifed 
icosahedron (solid) in figure~\ref{ico-design}, the 2057-atom cuboctahedron (dashed) and the 2057-atom Mackay 
icosahedron (dash-dot). The cut-off, at which $n_i^< (r_{\mbox{\scriptsize cut}}) = n_i$, is indicated by the 
vertical dashed line.}
\label{rho}
\end{figure}

We now wish to examine how the surface reconstruction improves the glue energy. It is instructive to look 
at the culmulative contribution to $\left< n_i \right>$, from pairs with $r_{ij}<r$ for atoms $i$ on the 
surface (defined here to be atoms with $n_i < 11$):
\begin{equation}
\label{culm}
\left< n_i^< (r)\right>_{\mbox{\scriptsize surf}} = {1 \over N_{\mbox{\scriptsize surf}}} \sum_{i \neq j, 
r_{ij}<r}^{\mbox{\scriptsize surf}} \rho (r_{ij}).
\end{equation}
This quantity, and the corresponding glue energy, are compared in figure~\ref{rho} for the 2057-atom 
cuboctahedron, Mackay icosahedron and resolidified icosahedron (note that the energy curve follows the shape
of the $\left< n_i^< (r)\right>_{\mbox{\scriptsize surf}}$-curve as $U$ is approximately linear away from its 
minimum). After the contribution to $\left< n_i \right>$ from the first shell ($r_{ij} < 4.25 \, \mbox{\AA}$ say), 
the Mackay icosahedron has the largest $\left< n_i^< \right>_{\mbox{\scriptsize surf}}$ value. This is to be expected 
as the Mackay 
icosahedron has only $\left\{ 111 \right\}$ facets at the surface. However, for the resolidified 
icosahedron $\left< n_i^< (r)\right>_{\mbox{\scriptsize surf}}$ becomes largest beyond $r=4.6 \, \mbox{\AA}$ 
where the contribution of the second-nearest neighbors begins to make an impact. Note that since $n_i^< 
(r_{\mbox{\scriptsize cut}}) = n_i$, it is clear from the figure that the resolidified icosahedra will 
have the lowest glue energy at the surface.

The relatively large contribution from next-nearest neighbors to $n_i$, arises due to the
small difference in energy between the $\left\{ 111 \right\}$ and $\left\{ 100 \right\}$ faces for lead
\cite{LOE92}. While an atom in a $\left\{ 100 \right\}$ face has fewer nearest-neighbors, it has more 
next-nearest neighbors than an atom in a $\left\{ 111 \right\}$ face. The glue energy reflects this 
small difference in energy by weighting the contributions from next-nearest neighbors more heavily 
than might otherwise be the case \cite{LOE92}. This would seem to be the key feature of the potential 
that stabilizes the novel surface reconstructions of the resolidified icosahedra. 

In the next section, we will explicitly construct new icosahedra from type B and C tetrahedra. This will 
enable us to compare these new icosahedra to fcc structures at cluster sizes where it is currently too 
expensive to conduct repeated resolidification trials.

\section{New Icosahedra}

Using the two types of tetrahedra identified in the resolidified icosahedra, we can construct $2^{20}$
icosahedra. However, many of the $2^{20}$ possible icosahedra can be identified after a rotation. In the 
appendix~\ref{appendix}, we show that there are only 17284 unique ways of constructing an icosahedron from 
the two types of tetrahedron.

This a large configuration space to search for the best arrangement of tetrahedra. However, a pair of clusters which 
are mirror images of one another will be energetically equivalent i.e. the pair will be chiral isomers. In fact, 
there are 1048 of the 17284 clusters which are invariant under reflections, leaving 8488 pairs of chiral isomers. 
This reduces the number of energetically distinct clusters to at most 9536. Further, the resolidification trials 
detailed in Hendy and Hall \cite{Hendy01} are a way of sampling this configuration space to discover low-energy clusters. 

To construct one of these new icosahedra from scratch, we begin with an $(n-2)$-shell Mackay icosahedra. At faces where
type-B tetrahedra are desired, add a further $(n-1)$-shell Mackay icosahedral face, and at faces where type-C tetrahedra 
are desired, add an $(n-1)$-shell anti-Mackay face. To complete the penultimate shell, Mackay icosahedral edges and 
vertices are added. Now to complete the outer shell, add an $n$-shell anti-Mackay face at faces where type-B tetrahedra 
are desired (the corresponding type-B tetrahedra are now layered as {\ldots}ABCA\underline{B}A), and an $(n+1)$-shell 
anti-Mackay face where 
type-C tetrahedra are desired (the corresponding type-C tetrahedra are now layered as {\ldots}ABC\underline{A}CB). 

The number of atoms in a n-shell icosahedra ($n>2$) constructed as above from $m$ type-C tetrahedra and $20-m$ type-B 
tetrahedra is then given by:
\begin{eqnarray}
\mbox{NewIco}(n,m) & = & \mbox{Ico}(n-2) \nonumber \\
& + & 30n-48 + m(n^2-n+1) \nonumber \\ 
& + & (20-m)(n^2-3n+3)
\label{newnumber}
\end{eqnarray}
The cluster from figure~\ref{ico-design}, which was the best structure produced in the trials shown in figure~\ref{solidtrial}, 
has seven type-C tetrahedra and thirteen type-B tetrahedra. With $m=7$, equation~\ref{newnumber} gives 
a sequence of 309, 565, 931, 1427, 2073, 2889, 3895, 5111, 8253 {\ldots} atoms in each closed-shell cluster. We note 
that a n-layer type-C 
tetrahedra contains $2(n-1)$ more atoms than the type-B tetrahedra and, hence a n-shell icosahedron with $m$ type-C 
tetrahedra has $2m(n-1)$ more atoms on the surface than an n-shell anti-Mackay icosahedron (which is constructed 
from 20 type-B tetrahedra). Thus, by increasing the proportion of type-C tetrahedra, the density of surface atoms 
is increased, relative to the anti-Mackay icosahedra, as seen in table~\ref{compare}.

In table~\ref{compare}, the energetics of a 2073-atom $m=7$ cluster, constructed using the design in figure~\ref{ico-design} 
and then relaxed, can be compared to those of a 2057-atom resolidified icosahedron. From the CNA-classification of the atoms 
in these clusters, it is clear that the two clusters are very similar in structure and energetics. Although, we did not 
explicitly include the extra \{100\}-atoms at the outer edges in the above construction, which appear as CNA signature 
C in the table, these arise as the constructed cluster relaxes. An interesting feature of the structure of this cluster 
is that it has a chiral isomer, as the $m=7$ pattern~\ref{ico-design} does not have a mirror symmetry. 

\begin{figure}
\resizebox{\columnwidth}{!}{\includegraphics{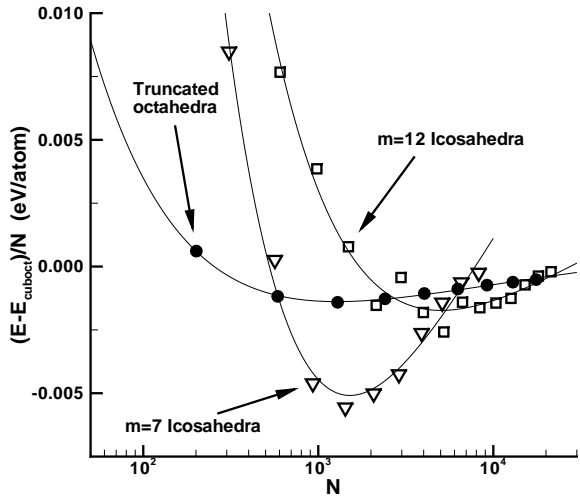}}
\caption{Energies of clusters versus size relative to the fit to the cuboctahedra sequence: $m=7$ icosahedra (open gradient symbol), 
$m=12$ icosahedra (open box) and truncated octahedra (filled circle).}
\label{icobest_energy}
\end{figure}

Figure~\ref{icobest_energy} shows the energies for a sequence of $m=7$ icosahedra, relative to the cuboctahedra 
sequence, with the first design given in figure~\ref{ico-design}. From figure~\ref{icobest_energy} it appears that 
this type of icosahedra is the lowest energy structure for sizes from 900-5000. The largest cluster of this design 
that lies above the interpolated fit to the truncated octahedra sequence is the 5111-atom cluster.    

We have found one other design which appears to be stable at larger sizes. This design is also shown in 
figure~\ref{ico-design} and has 12 type-C tetrahedra ($m=12$). The energies of this sequence are shown 
relative to the cuboctahedron fit in figure~\ref{icobest_energy}. 
The first energetically favored $m=12$ icosahedron occurs at a size of 5211 atoms. This sequence 
continues to give the most energetically favored clusters we have found until a cluster size of 
18097 atoms, where the truncated octahedron sequence appears to become more stable. Thus, this 
sequence gives the most energetically favored structures known from a size of 5211 to 15191 atoms.

\section{Discussion}

We have described a new family of icosahedral structures, which include
the anti-Mackay icosahedra. The family is characterized by a Mackay 
icosahedral core and a surface reconstruction that can extend into the 
two outer layers. They can be explicitly constructed from two
types of fcc tetrahedra: the first type of tetrahedra has an anti-Mackay
surface termination (type-B), and the second type of tetrahedra has a 
twin-plane fault in the third-outermost layer (type-C). The anti-Mackay
icosahedra is constructed entirely from type-B tetrahedra. Structures in 
this family of icosahedra were shown to be the lowest energy structures 
known for the lead glue potential (\ref{glue}) for certain sizes. This
study illustrates how new structural forms can arise when many-body effects 
are included in the interatomic potential.

Icosahedra constructed with a sufficient number of type-C tetrahedra 
possess a density of surface atoms comparable to that of the Mackay 
icosahedra, but without the low coordination atoms at the edges and
vertices of a Mackay icosahedron. Such a surface reconstruction was
shown to be favored by the glue-term in the interatomic potential
(\ref{glue}) over fcc structures largely due to the effects of 
next-nearest neighbors. We commented that the relatively large 
effect that the next-nearest neighbors have on the glue term is 
due to the small difference in energies between the $\{111\}$ and 
$\{100\}$ faces. This next-nearest neighbor contribution is also
the feature of the potential that favors fcc structures over Mackay
icosahedra at all sizes, as noted by Lim, Ong and Ercolessi \cite{LOE92} 
in their original study of lead clusters.

While we have not exhaustively searched all possible icosahedral structures 
that are part of this new family, we have identified two configurations of type-B 
and type-C tetrahedra that show particularly favorable energetics. The first
structure (which occurs in a pair of chiral isomers) has seven type-C tetrahedra 
and is the lowest energy structure known for the potential (\ref{glue}) over a size
range of 900-5000 atoms. The second structure (which is symmetric under reflection) 
has twelve type-C tetrahedra and is the lowest energy structure known for the potential 
(\ref{glue}) over a size range of 5000-18000 atoms. Thus, above 900 atoms, fcc structures 
are not favoured by (\ref{glue}) until at least cluster sizes of 18097 atoms, where the 
truncated octahedra appears to be favored. This is probably a conservative lower 
bound on the size where large fcc structures appear as global minima of the potential (\ref{glue}).

These results continue to emphasize the tendency of this potential to produce 
non-fcc structural forms. A recent global minimization of this potential for 
cluster sizes of up to 160 atoms, found that these clusters do not adopt fcc 
structures at any size in this range \cite{Doye02}. Further, it seems likely that no 
globally minimum fcc structure appears in the window between that study and the 
results presented here (i.e. between 160 and 900 atoms) although we have not explicitly 
demonstrated this here. Simulations of lead nanowires \cite{Gulersen98} have also shown the 
emergence of non-fcc structures. In all these cases, the non-fcc character is 
related to the small difference in energy between the $\{111\}$ and $\{100\}$ 
faces (although, disturbingly, it appears that the cut-off distance 
$r_{\mbox{\scriptsize}}$ for the glue energy also plays a role, at least for 
small clusters \cite{Doye02}).

We have focused here on icosahedral structures because they were produced in
the melting and freezing simulations of Hendy and Hall \cite{Hendy01}. While 
these structures were found to be lower in energy than other known structures, 
there is no guarantee that they are globally minimum. Indeed, simulations
of freezing often produce icosahedral structures preferentially, either for
thermodynamic or kinetic reasons, irrespective of whether these structures
are globally optimal (see for example Chushak and Bartell\cite{Chushak01}).
It is possible that the novel surface reconstructions seen here may also stabilize
decahedral forms, for example, which have not been seen in freezing simulations, 
but which may compete energetically with the icosahedral structures seen here.

\appendix*
\section{N-colored icosahedra}
\label{appendix}

We consider painting an icosahedron using up to N colors, where each face
can only be painted a single color. There are clearly $N^{20}$ ways to paint 
the icosahedron, but some of these painted icosahedra will simply be rotations 
of other painted icosahedra.

To count the number of unique painted icosahedra we use Burnside's theorem 
\cite{Burnside}. Let $X$ be the set of colorings of an icosahedron
($| X | =  N^{20}$) and $G$ be the rotational symmetry group of the 
icosahedron ( $| G |$ = 60, consisting of the identity, 15 $180^\circ$ rotations 
about edges, 20 $120^\circ$ rotations about faces, and 24 $72^\circ$ rotations about
vertices). Now for each $g \in G$, we define $X_g = \{ x \in X \, | \, gx=x\}$, where 
$gx$ is the new coloring obtained by rotating the coloring $x$ via the rotation $g$.

Now, Burnside's theorem gives the number of colorings, $C$, that are unique when acted 
upon by the finite group $G$:
\begin{equation}
C = {1 \over | G |} \sum_{g \in G} | X |_g
\end{equation}
Thus, we can determine $C$ by determining $| X |_g$  for each $g \in G$. In fact,
for each $g \in G$, $| X |_g = N^m$, where $m \leq 20$ is the number of orbits of 
faces under $g$. Determining the value of $m$ for each $g$ is straightforward, and 
can be done using a cardboard cut-out icosahedron (or counted on a computer).

Doing so, we arrive at the formula for $C$ for the N-colored icosahedron:
\begin{equation}
C = {1 \over 60} \left( N^{20}+24 N^{4}+20 N^{8}+15 N^{10} \right)
\end{equation}
Thus for $N=2$, the number of 2-colored icosahedra is 17824.
 
We can also consider N-colored icosahedra under reflections as well as rotations. The size 
of the symmetry group is now doubled as we add a generator of reflections 
to the group: $| G |$ = 120. The formula for the number of N-colored icosahedra 
unique under rotations and reflections is:
\begin{eqnarray}
\nonumber C & = & {1 \over 120} \left( N^{20}+N^{10}+24(N^2+N^4) \right. \\ 
\label{chiral} & + & \left. 20(N^8+N^4)+15 (N^{12}+N^{10}) \right) 
\end{eqnarray}
Thus, if we identify mirror images, formula (\ref{chiral}) applies, and hence the number 
of 2-colored icosahedra is 9436 ($N=2$). Of the 17824 2-colored icosahedra, 1048 are invariant 
under reflection symmetry, while the remaining 16776 2-colored icosahedra come in 8388 pairs
of mirror images. 

\begin{acknowledgments}
The authors wish to thank Peter McGavin for his help with parts of the
appendix material. J. P. K. D. is grateful to the Royal Society for the award
of a University Research Fellowship. S. C. H. would like to acknowledge the 
support of the ISAT linkages fund administered by the Royal Society of New Zealand.
\end{acknowledgments}

\end{document}